\documentclass[twocolumn]{aastex631}
\usepackage{amsmath}
\usepackage[utf8]{inputenc}
\usepackage[T1]{fontenc}

\newcommand{\BIBORDERINGTAG}[1]{}

\shorttitle{Continuously Resolved Starburst Galaxy and Quasar Spectrum}

\begin{document}

\title{A Hot DOG Forged in FIRE: Nuclear and Starburst Spectral Decomposition of a Luminous Infrared Galaxy Simulation with a Resolved Dust Torus}

\author[0009-0002-8417-4480]{Jaeden Bardati}
\affiliation{TAPIR, Mailcode 350-17, California Institute of Technology, Pasadena, CA 91125, USA}
%\email{jbardati@caltech.edu}
\correspondingauthor{Jaeden Bardati}
\email{jbardati@caltech.edu}

\author[0000-0003-3729-1684]{Philip F. Hopkins}
\affiliation{TAPIR, Mailcode 350-17, California Institute of Technology, Pasadena, CA 91125, USA}
%\email{phopkins@caltech.edu}

\author[0000-0002-4900-6628]{Claude-Andr\'{e} Faucher-Gigu\`{e}re}
\affiliation{CIERA \& Department of Physics and Astronomy, Northwestern University,
1800 Sherman Ave, Evanston, IL 60201, USA}

% TO DO:
% - Add dust attenuation to fig 1 ? make a zoom-in plot instead? (1 Mpc -> 100 kpc --> ... --> 1 pc images) + change figure caption?
% - "near- to mid-IR cutoff" --> "near/mid-IR suppression"

% Done since submit: 
% - added comments/citation from Tadaki
% - 

%%%%%%%%%%%%%%%%%%%%%%%%%%%%%%%%%%%%%%%%%%%%%%%%%%%%%%%%%%%%%%%%%%%%%%%%%%%%%%
\begin{abstract}
Ultraluminous infrared galaxies are powered by a combination of rapid star formation and active galactic nucleus (AGN) emission, but their relative importance is not always observationally clear. 
We study the galactic continuum spectrum of a cosmologically simulated $\sim 4 \times 10^{10} M_\odot$ stellar mass starburst galaxy at redshift $z\sim 4.4$ that refines down to resolve beyond the dust sublimation boundary of its super-Eddington-accreting $\sim 10^7 M_\odot$ supermassive black hole.
We find that this system resembles the rare class of hot dust-obscured galaxy (Hot DOG), with a roughly flat (in $\nu F_\nu$) IR emission spectrum that sharply drops off at wavelengths $\lesssim 5~\mu\mathrm{m}$. 
Our system also matches with the observational properties of many Hot DOGs, including undergoing multiple galaxy mergers and being the most massive galaxy within a dense cosmological environment.
The distinctive Hot DOG spectral shape in our system is caused by AGN-heated mid-IR warm dust, predominately starburst-heated far-IR cold dust, and a steep near- to mid-IR cutoff caused by strong absorption in the dense ISM of the galactic nucleus, rather than the dust torus itself. 
This system is lower luminosity ($L_\mathrm{IR} \sim 2 \times 10^{12} L_\odot$) than those detected by the WISE survey at similar redshifts, but will be a prime target for future far-IR surveys such as PRIMA.
Our results show that Hot DOGs can naturally result as a transitional phase during rapid AGN accretion, but before significant AGN-driven outflows clear optically thin paths. 
\end{abstract}

% https://astrothesaurus.org/concept-select/

%%%%%%%%%%%%%%%%%%%%%%%%%%%%%%%%%%%%%%%%%%%%%%%%%%%%%%%%%%%%%%%%%%%%%%%%%
\section{Introduction} \label{sec:intro}

Ultraluminous infrared galaxies (ULIRGs; defined as galaxies with IR luminosity $L_{\mathrm{8-1000}\mu\mathrm{m}} > 10^{12} L_\odot$) are among the most luminous sources in the universe, and are thought to dominate the cosmic infrared luminosity at redshifts $1 \lesssim z \lesssim 4$ \citep{Hopkins_2006, Hopkins_2008, Magnelli_2013, Algera_2023, Koprowski_2025}. 
The intense IR luminosity in these galaxies is thought to be due to infrared re-emission of dust-absorbed optical/UV emission from a period of extreme rapid star formation \citep[starburst;][]{Rowan_Robinson_2000, Casey_2014}, an actively accreting supermassive black hole \citep[SMBH;][]{Weedman_2012, Tsai_2018}, or some combination of the two \citep[][]{Sanders_1996, Nardini_2008, Veilleux_2009, Kirkpatrick_2015}. 

Observational evidence \citep[see review by][]{Lonsdale_2006} and idealized simulations \citep[e.g.][]{DiMatteo_2005, Hopkins_2005, Hopkins_2008, Hopkins_2010} suggest that ULIRGs at $z\sim 0$ result from major gas-rich mergers and represent a key stage in galaxy-SMBH co-evolution \citep[see review by][]{Kormendy_2013}. Such galaxy mergers trigger both a period of intense star formation and dusty inflows that feed the central SMBH. These systems are then thought to undergo ``buried" active galactic nuclei (AGN) phases in which the SMBH accretion disk is heavily obscured by the surrounding dust and after which SMBH feedback energy can cause an outflow that expels the surrounding obscuring medium. This feedback can reveal an unobscured quasar \citep{Hopkins_2016} and can quench the galaxy, shutting off star formation \citep{Hopkins_2006}. 

However, it is not always observationally easy to disentangle the starburst and AGN components of the IR emission. In some systems, particularly strong starbursts can produce sufficient thermal dust emission from reprocessed starlight in ULIRGs to mask any possible AGN emission under the galactic continuum \citep{Chakrabarti_2007}, such as with some submillimeter galaxies \citep[SMGs;][]{Blain_2002}. In other cases, the AGN may contribute strongly enough to the ISM dust heating to cause an enhanced mid-IR spectrum and produce so-called ``warm" ULIRGs \citep{Sanders_1988, Armus_2007}. There are also much rarer cases that are thought to be almost entirely AGN dominated such as hot dust obscured galaxies \citep[Hot DOGs;][]{Eisenhardt_2012}. These galaxies were discovered by the Wide-field Infrared Survey Explorer (WISE) as W1 ($\lambda_\mathrm{eff} \sim 3.4 \mu\mathrm{m}$) and W2 ($\lambda_\mathrm{eff} \sim 4.6 \mu\mathrm{m}$) band dropouts, characterized by power-law IR spectral slopes that cut off around $\sim 3-10~\mu\mathrm{m}$ and are among the most luminous galaxies currently known \citep{Wu_2012, Assef_2015}. The WISE Hot DOGs typically have redshifts near cosmic noon ($1 \lesssim z \lesssim 4$) and have Atacama Large Millimeter Array (ALMA) follow-up imaging that indicate that these systems are often found undergoing merger events \citep{Fan_2016}, suggesting that they may also be highly star-forming. However, flux limitations of current IR surveys make it hard to search for low-luminosity Hot DOGs which may have more prevalent starburst contributions to the spectral shape. Indeed, since many observations of rare ULIRGs are flux-limited or poorly spatially resolved, magnetohydrodynamics (MHD) simulations are vital to understanding the underlying physics and evolution of ULIRGs.

Although there have been many predictions of ULIRG galaxy spectra at the AGN scale, using idealized simulations or observationally-informed modeling of dust tori \citep[e.g.,][]{Fritz_2006, Honig_2006, Nenkova_2008, Stalevski_2012}, or at the galactic scale, using simulations of galaxy evolution that either scale empirical spectra \citep[e.g.,][]{Hopkins_2010b} or impose sub-grid prescriptions for the dust torus \citep[e.g.,][]{Narayanan_2010_pldog}, they generally do not self-consistently resolve both the galaxy and the AGN components simultaneously. Yet, simulating both regimes and their boundary simultaneously is key to understanding the heating source of ULIRGs and producing realistic synthetic quasar spectra. This is particularly important in the near to mid-IR produced by the galactic nucleus, where the AGN-heated hot dust must be resolved in conjunction with the stellar-heating from the central starburst, and which produces a significant amount of the difference between certain ULIRG galaxy subtypes. The challenge in simulating both galaxy and dust torus regimes lies in having sufficient physics and numerical techniques to self-consistently zoom-in to simulate and resolve the $\sim 12$ orders of magnitude separating galactic and supermassive black hole scales. Recently, this was achieved in \citet[][FORGE'd in FIRE]{Hopkins_2024a} which zoomed-in to around $\sim 300 R_s$ from the $M_\mathrm{BH}\sim 1.3\times 10^7 M_\odot$ central SMBH of a starburst galaxy merger at redshift $z = 4.4$. A unique finding of this simulation was the generation of a turbulently-supported accretion disk dominated by torodial magnetic fields that formed naturally as a result of advection from the nuclear ISM \citep{Hopkins_2024b}. In \citet{Bardati_2026}, we presented the infrared spectral emission from the dust torus simulated in this FORGE'd in FIRE simulation and showed that the AGN system is buried and IR-bright. The dust torus is primarily composed of the outer accretion disk and an inflowing cold dust stream feeding accretion from the bulk surrounding ISM complex. This inflowing dusty stream causes significant anisotropy in the near- to mid-IR emission. Since the starburst in this system has a combined energy of $\sim 2 \times 10^{45}~\mathrm{erg~s^{-1}}$, which is comparable to the luminosity of the supermassive black hole $L_\mathrm{AGN} \sim 5 \times 10^{45}~\mathrm{erg~s^{-1}}$, there should be similar amounts of stellar and AGN dust heating.

In this paper, we extend the dust torus radiative analysis from \citet{Bardati_2026} to include galactic contributions, producing the UV/optical to infrared spectra of a simulated ULIRG with resolved dust from the AGN sublimation boundary to galactic scales. In Section \ref{sec:simulations}, we describe our FORGE'd in FIRE simulation and our post-processing radiative transfer simulations with SKIRT. We describe and discuss our results in Section \ref{sec:results} and summarize our conclusions in Section \ref{sec:conclusion}.

 \section{Simulations}\label{sec:simulations}

In this section, we briefly describe our FORGE'd in FIRE zoom-in simulation and our post-processing radiative transfer simulations with SKIRT. Since these methods each are described extensively in their own papers \citep{Hopkins_2024a, Bardati_2026}, we only focus on the most salient points here. 
 
\subsection{FORGE'd in FIRE}

\begin{figure*}[t]
    \centering
    \includegraphics[width=\textwidth]{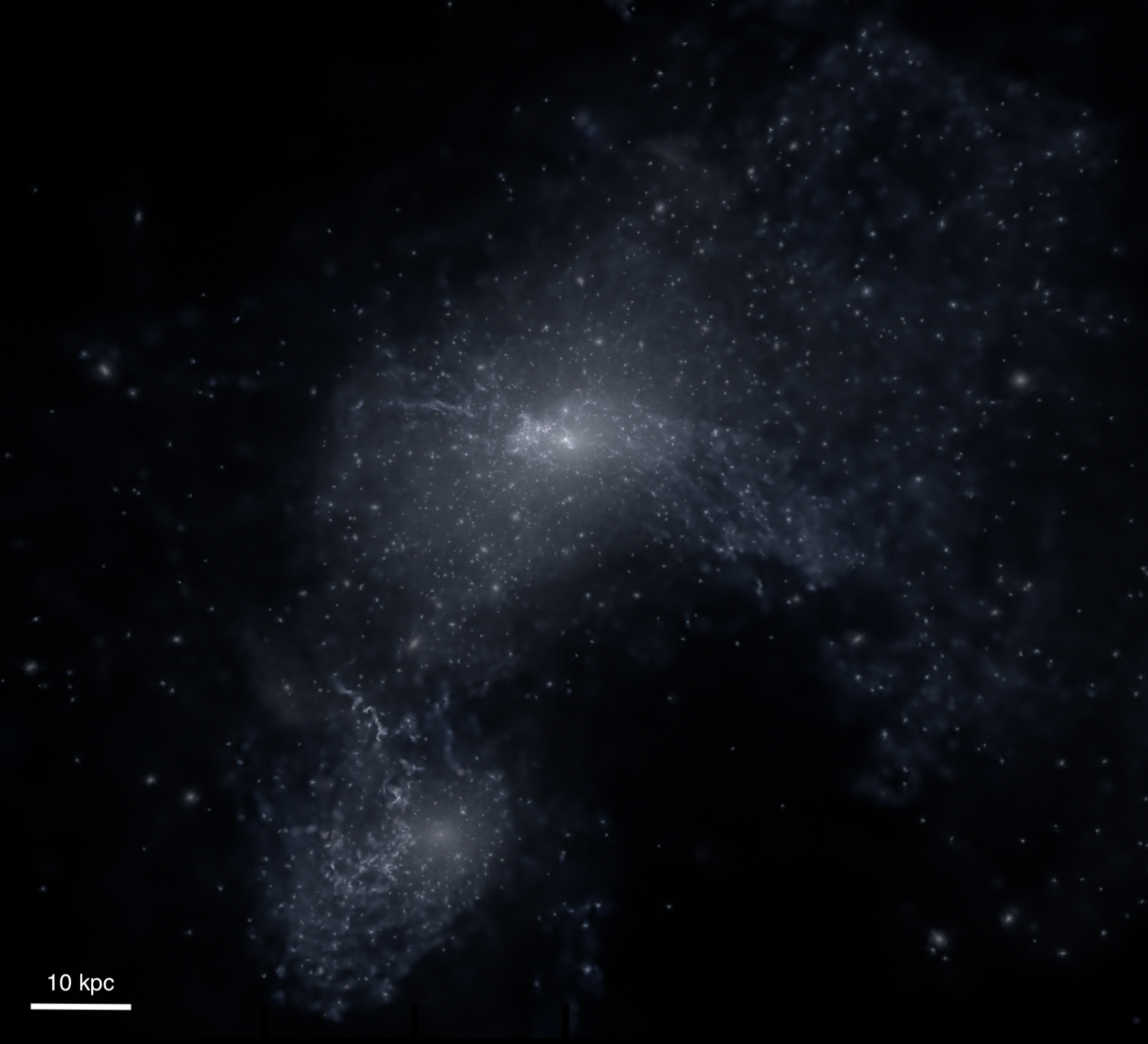}
    \caption{Spatially resolved optical stellar emission (without dust attenuation) of the FORGE'd in FIRE galaxy in rest-frame optical bands. The system is clearly undergoing a merger, has a highly clumpy morphology and is undergoing a starburst causing the optical/UV stellar emission to appear blue. This system is the most massive galaxy in its dense cosmological environment at redshift $z\sim4.4$. 
    }
    \label{fig:galaxy_image}
\end{figure*}

We use a FORGE'd in FIRE radiation-magnetohydrodynamics (MHD) zoom-in simulation run using the GIZMO meshless finite-mass (MFM) code \citep{Hopkins_2015}. The simulation combines the large-scale FIRE-3 \citep{Hopkins_2023FIRE3} and small-scale STARFORGE \citep{Grudic_2021} physics to self-consistently zoom-in to a $M_\mathrm{BH} \approx 1.3 \times 10^7 M_\odot$ super-Eddington accreting black hole. The simulation starts with a 100 Mpc per-side box at $z\sim 100$ and evolves using standard $\Lambda$CDM cosmology and galaxy formation physics. The simulation continues until $z \sim 4.4$ and refines to resolve a $\sim 100$ kpc galaxy merger event that is undergoing a starburst period. The SMBH zoomed into is in the center of the most massive galaxy of the ``A1"/``m12z4" system previously studied in \citet{Feldmann_2016, Feldmann_2017, Oklopvcic_2017, Angles_Alcazar_2017, Hopkins_2020b, Ma_2021, Wellons_2023}. Characteristic of early universe galaxies, the galaxy is morphologically clumpy (see Figure \ref{fig:galaxy_image}), has a young stellar population (mean stellar formation age $t_\mathrm{form} \sim 400~\mathrm{Myr}$), and has low metallicity ($Z \sim 0.2 Z_\odot$). The simulation continues to resolve around the central SMBH until it reaches $\sim 300$ Schwarzschild radii separation, well below the $\sim 0.25$ pc dust sublimation region \citep{Bardati_2026}. The simulation includes numerous physics implementations, including full self-gravity with high-order Hermite integrators for accurate multiple orbits \citep{Grudic_2020, Grudic_2021b, Hopkins_2023}, non-ideal magnetohydrodynamics \citep{Su_2017, Hopkins_2017}, multi-band M1 radiation-hydrodynamics \citep{Hopkins_2019, Hopkins_2020}, dust, molecular, atomic, metal-line, and ionized gas thermochemistry fully-coupled to radiation \citep{Colbrook_2017, Grudic_2021, Choban_2022}, SMBH accretion and seeding \citep{Hopkins_2016, Shi_2023, Wellons_2023}, and both stellar population and individual resolved (proto)star evolution and feedback \citep{Grudic_2021, Grudic_2022, Hopkins_2023FIRE3}. Although this simulation has since been run to the innermost stable circular orbit (ISCO) of the black hole \citep{Hopkins_2025}, we use the version presented in \citep{Hopkins_2024a} that has been run to $\sim 300 R_s$, resolving well past the dust torus and end of star formation.

Since the simulation covers $\sim 12$ orders of magnitude in length scale, resolving down to individual star formation, stars are treated with two distinct particle types. Low mass resolution ($M > 1 M_\odot$) stars are modeled as stellar population particles and handled by FIRE physics \citep{Hopkins_2023FIRE3}, with stellar population formation using the entire \citet{Kroupa_2001} initial mass function (IMF) and feedback via IMF-sampled main-sequence tracks including radiation, mass-loss, supernovae, and cosmic rays \citep{Su_2018}. At high resolutions ($M \lesssim 1 M_\odot$), stars are treated as individual stars and handled by STARFORGE physics \citep{STARFORGE} with resolved Larson cores during formation, evolving along proto-stellar/main-sequence single-star tracks with feedback from radiation, jets, mass-loss on the surface and end-of-life explosions. Both particles are used throughout the simulation as it is refined to allow for a smooth transition between the resolution scales without explicitly or abruptly changing physics models at any point. To further help avoid any possible transition issues, the simulation is rapidly refined through the mass resolution transition zone where $10^{-2}~M_\odot< \delta M < 10^2~M_\odot$ \citep[see][]{Hopkins_2024a}. We refer the reader to \citet{Hopkins_2024a} for more specific details on the physics implementation and system details.

 \subsection{Radiative Transfer with SKIRT}

In post-processing, we run the 3D Monte Carlo dust radiative transfer code SKIRT \citep{SKIRT_2020} on our FORGE'd in FIRE simulation to calculate the emergent spectra. %SKIRT features support for custom emission point or extended sources, stellar population synthesis, multiple anisotropic scattering absorption and re-emission by dust and free electrons, Doppler shifting, polarization from dust scattering or from magnetically-aligned spheroidal dust grains, and non-LTE medium state and radiation field iteration including dust destruction and self-absorption. 
We broadly follow the configurations used in \citet{Bardati_2026}, but add the contribution of larger scales ($r > 10$ pc from the SMBH) and include stellar contributions which become relevant at $r \gtrsim 100$ pc. Specifically, our primary radiation sources are composed of a power-law accretion disk radiation source for the system \citep[described in detail in][]{Bardati_2026} with a simulation-informed luminosity of $L_\mathrm{AGN} \approx 3L_\mathrm{Edd} \approx 5 \times 10^{45}~\mathrm{erg/s}$, as well as both types of star particles in FORGE'd in FIRE. We note that since this system is entirely optically thick to the accretion disk, the exact shape of the accretion disk matter little to the resulting optical-IR spectrum since they will be reprocessed by the dust anyway \citep[see discussion in][]{Bardati_2026}. These primary radiation sources are reprocessed by the surrounding medium through Compton scattering off free electrons and scattering, absorption and re-emission by \citet{Weingartner_2001} Milky Way-like dust. The state of the dust medium (temperature and sublimation state) is iterated over until radiative equilibrium is reached. A dust cell is considered sublimated and its density (and thus opacity and emissivity) is set to zero when, in a given iteration, the cell mean temperature exceeds the sublimation temperature (with d$T_\mathrm{sub,~silicate} = 1200\mathrm{~K}$ and $T_\mathrm{sub,~graphite} = 1800\mathrm{~K}$).

Since it is not computationally feasible to run the Monte Carlo algorithm on the whole domain, we start with the mean spectrum output calculated using material within a 1 pc box \citep[namely, {the mean of the $1000$ isotropic sightlines} from][]{Bardati_2026} {and} input it as a {point} source in a {similarly configured} radiative simulation with a box containing the next 10 pc of material{. We then, take the mean of $1000$ isotropic sightlines in this simulation and input it as a source in a simulation with the next 100 pc of material. We continue this iterative process for each order of magnitude a total of 5 times to} reach a {maximum} box size of 100 kpc which fully encapsulates the central galaxy and merger complex. {This is decent approximation given our galaxy morphology at large scales and given the more isotropic, optically thick layer of ISM dust surrounding the torus that thermalizes the AGN emission at small scales. We confirm that this is the case by comparing this method with a simulation using all material $r<10~\mathrm{pc}$.}

New to this paper is the contribution of individual stars and proto-stars from STARFORGE and the contribution of stellar population particles from FIRE. To match the STARFORGE criteria \citep{STARFORGE}, we model the individually resolved stars as blackbodies with effective temperature and radius given by the simulation particle, which are mostly unresolved at the scales we study here. {While, in principle, the resolved star particles could deviate significantly from from a blackbody (such as the UV-deficit of Wolf-Rayet stars, which could overestimate the amount of stellar-heating at small radii), the total luminosity of these particles is small enough in comparison to the AGN-heating or stellar-heating for this not to be significant effect on the dust-heating and emergent spectra.} We model FIRE particles as simple stellar populations (SSPs) using the STARBURST99 \citep{Leitherer_2014} stellar models, using a \citet{Kroupa_2001} IMF, consistent with the FIRE-3 physics \citep{Hopkins_2023FIRE3}. In order to avoid unrealistic granularity or over-smoothing, we also give each SSP particle a smoothing length corresponding to the distance to its 32nd nearest neighbour SSP particle \citep[similar to e.g.][]{Trayford_2017}. {Despite treating the FIRE particles as SSPs in post-processing, at the radii where deviation from an SSP could be relevant, AGN-heating dominates the dust temperatures and have significantly lower total luminosities compared to the surrounding galactic population of stars.}

%%%%%%%%%%%%%%%%%%%%%%%%%%%
\section{Results \& Discussion} \label{sec:results}

In this section, we describe and discuss our results including comparing our total galaxy and quasar spectrum to various observed ULIRG galaxy types, the decomposition of starburst and AGN components, the cause of the hot dust obscured galaxy (Hot DOG) shape in our system, and a discussion on such low-luminosity Hot DOGs and their potential detection with proposed instruments.

\subsection{Hot Dust Obscured Galaxy Spectrum}

\begin{figure*}[t]
    \centering
    \includegraphics[width=\textwidth]{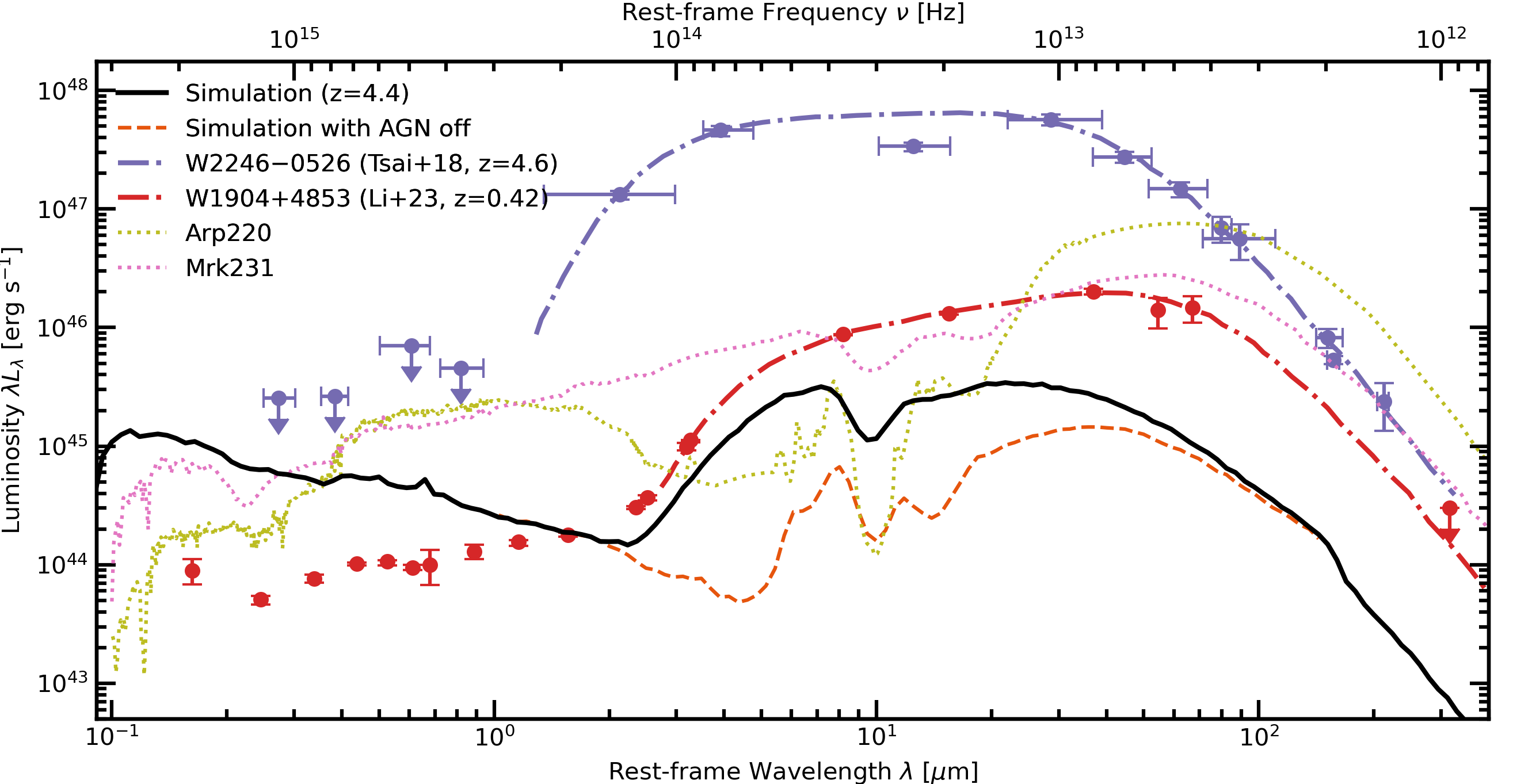}
    \caption{Rest-frame sightline-mean spectrum of the galactic total emission ($r \leq 100$ kpc) when including both the fully resolved AGN-heated and starburst-heated dust (solid black line) and compare it to the rest-frame spectra of various observed ultraluminous infrared galaxies (ULIRGs). We show the most luminous hot dust obscured galaxy (Hot DOG) W2246-0526 spectral measurements from \citet[][in purple]{Tsai_2018} and the low-redshift Hot DOG W1904+4853 spectrum from \citet[][in red]{Li_2023}, with best fits in dot-dashed lines. We also show local ULIRGs Arp220 (yellow) and Mrk231 (pink) in dotted lines \citep{Polletta_2007}. The dust continuum component of our spectrum is mostly flat in $\nu L_\nu$ from $\sim 5 \mu\mathrm{m}$ to $\sim 50 \mu\mathrm{m}$ and closely resembles the shape of a Hot DOG. Our simulation is lower luminosity than most observed Hot DOGs (similar to low-redshift Hot DOGs) at similar redshift ($z\sim 4$).}
    \label{fig:spectral_comparison} %%% ADD SIGHTLINE VARIATION FOR OPTICAL/UV ???!!??
\end{figure*} 

\begin{figure}[h!t]
    \centering
    \includegraphics[width=0.485\textwidth]{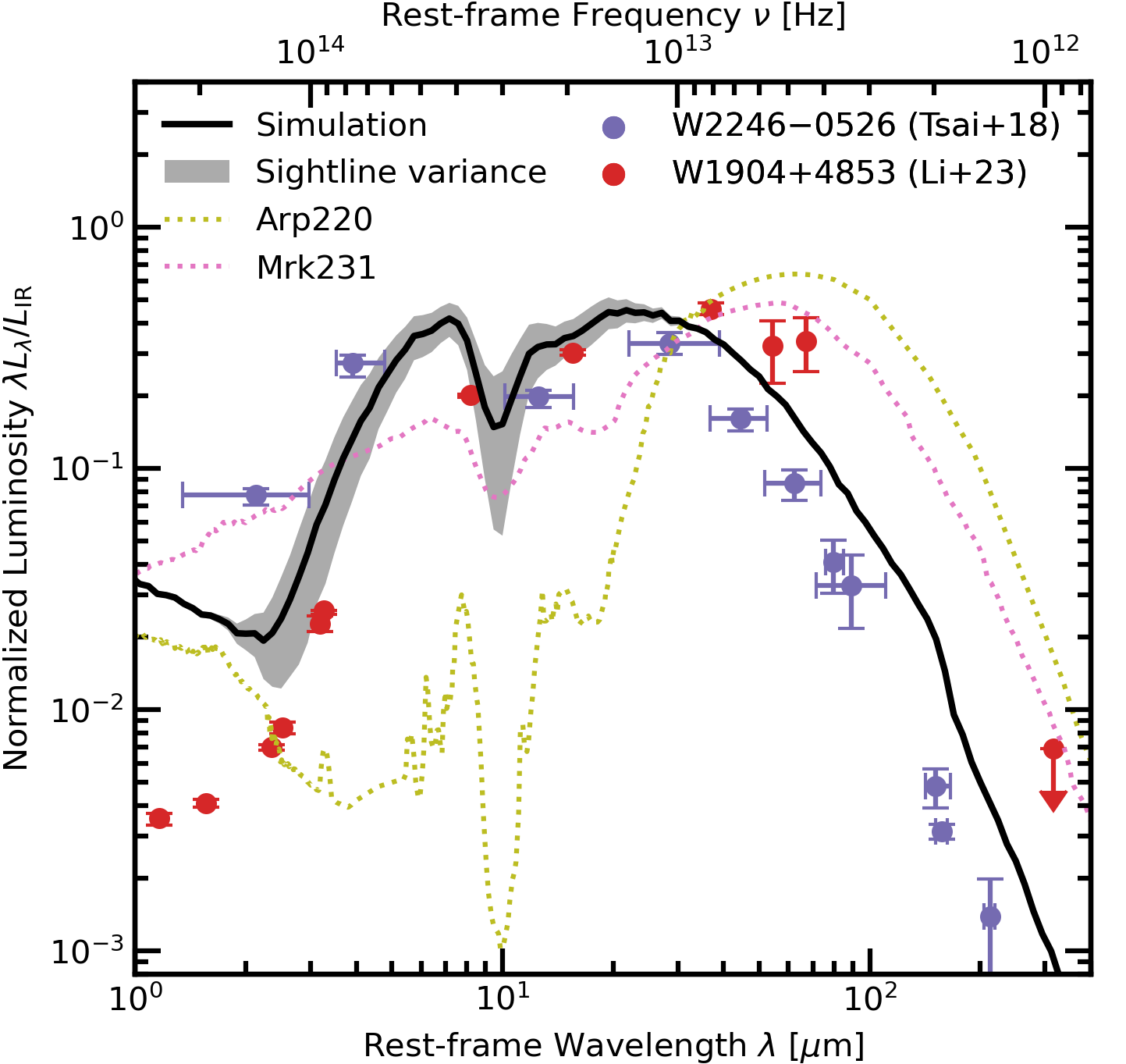}
    \caption{Rest-frame sightline-mean spectrum of the galactic total emission ($r \leq 100$ kpc) as shown in Figure \ref{fig:spectral_comparison}, but normalized in IR luminosity for a better IR spectral shape comparison. We also include a conservative estimate for the $1\sigma$ sightline variance in our spectra (assuming most of the sightline IR variance comes from $r < 100$ pc, where most of the dust self-absorption occurs). Our spectrum is broadly consistent with W2246-0526 and the low-redshift W1904+4853 spectrum in the mid-IR, but is likely different in the near- and far-IR due to changes in the surrounding host galaxy and relative stellar heating contribution. }
    \label{fig:spectral_comparison_renorm}
\end{figure}

We show the broadband rest-frame spectrum resulting from our post-processing of out FORGE'd in FIRE simulation in Figure \ref{fig:spectral_comparison}, accounting for AGN- and starburst-heated dust at scales ranging from the sublimation region at the inner torus boundary ($r \sim 0.1$ pc) to the interstellar medium ($r \sim 100$ kpc). We also compare it next to the spectral distribution of some local ULIRGs \citep[Arp 220 and Mrk 231;][]{Polletta_2007} and some distant hot dust obscured galaxies \citep[WISE J224607.57-052635.0 aka W2246-0526 and WISE J190445.04+485308.9 aka W1904+4853;][]{Tsai_2018, Li_2023}. Broadly, our system exhibits a flat (in $\nu F_\nu$) infrared continuum spectrum that is hotter than a typical ULIRG, but falls off sharply in the near infrared. The UV/optical component is steep and, as will be discussed in more detail later (Section \ref{sec:spectral_decomposition}), is dominated by stellar emission from a galactic starburst of young, low-metallicity ($Z \sim 0.2 Z_\odot$) stars, characteristic of galaxies at high-redshifts. 

Among the ULIRG examples shown in Figure \ref{fig:spectral_comparison}, our system most resembles the spectral shape of a hot dust obscured galaxy (Hot DOG). Hot DOGs are a population of rare high-redshift ($1 \lesssim z \lesssim 4$) luminous infrared galaxies characterized by an unusually hot dust component thought to be associated to an AGN spectrum that cuts sharply around a few microns \citep{Wu_2012, Assef_2015}. This distinguishes this class of ULIRGs from a typical power-law DOG (PL DOG) such as Mrk231, that extends as mostly featureless power-law from the optical to mid-IR. PL DOGs otherwise share many similarities to Hot DOGs such as how they are primarily powered by quasars and are thought to be a transitionary galactic state in galaxy mergers between a starburst galaxy and a typical type 1 quasar \citep{Narayanan_2010, Fan_2016, Diaz_Santos_2018}. Our system is also very distinct from typical luminous submillimeter galaxies (SMGs) such as Arp220, which are primarily powered by stellar emission from a starburst, have prominent cold dust continua and strong polycyclic aromatic hydrocarbon (PAH) features, yet also represent a key stage in galaxy mergers \citep{Chapman_2005, Miettinen_2017}. 

\begin{figure*}
    \centering
    \includegraphics[width=\linewidth]{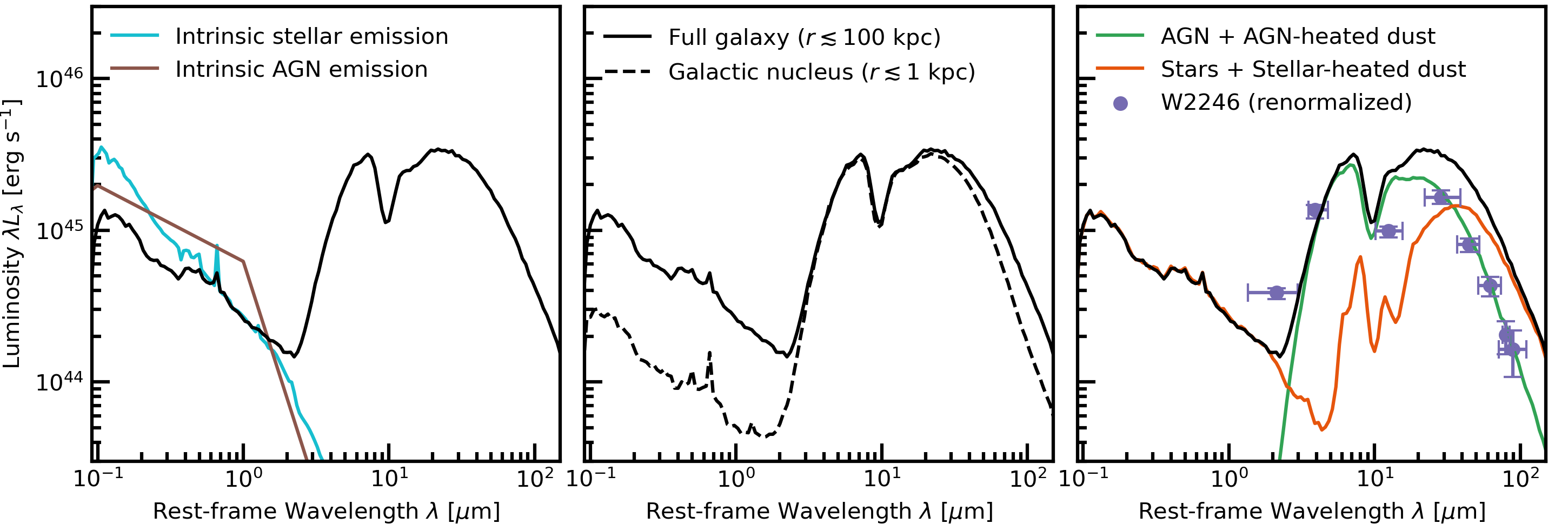}
    \caption{\textit{Left panel:} Intrinsic spectra vs dust reprocessed spectrum. We compare the intrinsic (non-dust attenuated) spectrum to the full galaxy spectrum (as in Figure \ref{fig:spectral_comparison} and \ref{fig:spectral_comparison_renorm}) in solid black. Most of the emergent (stellar) UV/optical emission is only slightly reddened. We note that our input AGN spectrum \citep[described in][]{Bardati_2026} does not include any dust torus component since we fully resolve the dust torus in this simulation. \textit{Middle panel:} Full galaxy spectrum vs galactic nucleus spectrum. We compare the emergent spectrum from the $r \lesssim 100$ kpc ($\sim 14\farcs7$ at $z= 4.4$) galaxy (solid black) to the spectrum emergent spectrum from the galactic nucleus at $r \lesssim 1$ kpc  ($\sim 147$ mas, in dotted black). The majority of the mid-IR emission is from the galactic nucleus, whereas the UV/optical and far-IR emission is primarily produced outside the nucleus \textit{Right panel:} Spectral decomposition of the stellar- vs AGN-heated dust. We compare the stellar (dust attenuated) emission and stellar-heated dust (in orange) to the {emergent} AGN {(disk and torus)} emission and AGN-heated dust (in green). The mid-IR is dominated by the AGN-powered nucleus dust and the far-IR by the starburst-powered cold extended galactic dust, reaching roughly equal contributions around 40 $\mu\mathrm{m}$. We also note that renormalizing the W2246-0526 spectrum to $L_\mathrm{AGN} = 5 \times 10^{45} \mathrm{~erg~s}^{-1}$ (in purple), lines up surprisingly well with our AGN-heating spectrum, consistent with the understanding of W2246-0526 being AGN-heating dominated.}
    \label{fig:spectral_decomposition}
\end{figure*}

We compare our galaxy to the rest-frame spectra of the most luminous and well studied Hot DOG W2446-0526 (at redshift $z\sim 4.6$) and of the low-redshift Hot DOG W1904+4853 in Figure \ref{fig:spectral_comparison} (without normalization) and Figure \ref{fig:spectral_comparison_renorm} (with normalization). The flat power law IR shape with near-IR cutoff characteristic of Hot DOGs is present in all three spectra and are all broadly consistent with each other in the mid-IR. We do note, however, that the cutoff potentially occurs at different temperatures.  It is unclear exactly what could cause the variation in near-IR cutoff, but we speculate in Section \ref{sec:hotdog_shape} that this may be due to varying degrees of galactic nucleus ISM dust self-absorption.
The spectral shapes diverge in the far-IR and UV/optical, but this is likely just a difference in the surrounding galaxy and stellar formation rates relative to the black hole luminosity. As we illustrate later, the W2446-0526 spectrum in particular is strikingly consistent with the AGN{-heated} portion of {our simulated spectrum} (see Figure \ref{fig:spectral_decomposition}), which is consistent with it being mostly AGN dominated. On the other hand, W1904+4853 may have proportionally more star formation, causing it to have a more prominent far-IR tail (Figure \ref{fig:spectral_comparison_renorm}).

W2446-0526 and other high-redshift ($z\sim 2-4$) Hot DOGs are very luminous due to the flux-limited constraints of the WISE survey, however relatively nearby Hot DOGs such as W1904+4853 have been discovered at redshifts $z < 0.5$ with WISE and Herschel archival data \citep[e.g.][]{Li_2023}. These low-redshift, lower luminosity Hot DOGs have been shown to have lower BH masses and lie closer to the relation between galaxy stellar mass and black hole mass \citep{Li_2025}. Our system seems to share properties of both systems. Namely, it is similar redshift to high-redshift Hot DOGs ($z=4.4$), but has similar luminosities to low-redshift Hot DOGs. Thus, we might expect our system to have a lower metallicity {and, by construction,} younger population of stars when compared to low-redshift Hot DOGs. Indeed, the stellar-dominated UV/optical component of our system is significantly steeper than W1904+4853 (or Arp220 and Mrk231), typical of young, low-metallicity (here, $Z \sim 0.2 Z_\odot$) galaxies found at high-redshifts. {Of course, it is important to note that differences between dust content and geometry between the systems can also play a significant role in the UV/optical spectrum, and it is possible the dust is more centrally concentrated in our particular system.} It is worth noting that this steep UV/optical spectral index coupled with the hot IR emission could lead to this galaxy being classified as a type 1 AGN in photometric surveys with low enough spatial resolution. Of course, spectroscopic follow-up would later reveal only stellar lines and higher spatial resolution photometry would resolve the extended stellar emission. 

Aside from its spectrum, our galaxy system also shares many similarities to observed Hot DOGs. Most notably, it is undergoing hierarchical mergers (see Figure \ref{fig:galaxy_image}) which is consistent with the majority of Hot DOG follow-ups with Hubble Space Telescope near-IR imaging \citep{Fan_2016} and ALMA sub-millimeter imaging \citep{Diaz_Santos_2018, Fernandez_Aranda_2025}. Our system is particularly chaotic and clumpy, characteristic of bursty starforming galaxies in the early universe \citep{Hopkins_2012, Feldmann_2016, Oklocvic_2017, Gurvich_2023, Hopkins_2023b, Sun_2023}. In particular, our system is the most massive galaxy of a dense cosmological environment which is also consistent with observational findings for Hot DOGs \citep{Assef_2015, Ginolfi_2022, Zewdie_2023, Luo_2024, Zewdie_2025}. The active central SMBHs of Hot DOGs are thought to highly accreting at rates near- or super-Eddington \citep{Ricci_2017, Li_2024} that are obscured by enough gas and dust to make the system Compton-thick in the X-ray \citep{Zappacosta_2018, Zou_2025}, which is also consistent with this system \citep{Bardati_2026}. Our system has star formation rate SFR $\sim 10^2~M_\odot/\mathrm{yr}$ a stellar mass of $M_* \sim 4 \times 10^{10} M_\odot$, black hole mass of $M_\mathrm{BH} \sim 1.3 \times 10^7 M_\odot$, dust mass of $M_\mathrm{dust} \sim 10^8 M_\odot$ (nucleus dust of $M_\mathrm{dust, r<1kpc} \sim 5 \times 10^6 M_\odot$) and a star formation to AGN IR luminosity ratio of $L_\mathrm{IR, SF}/L_\mathrm{IR, AGN} \sim 0.7$. Thus, our system lies on or slightly above the galaxy SFR-$M_*$ and $M_\mathrm{BH}$-$M_*$ relations, consistent with many Hot DOGs \citep{Fan_2018, Li_2023, Li_2025}. The galaxy kinematics of our system are very chaotic, consistent with some Hot DOGs, though they have been shown to have a wide range of kinematics observationally \citep{Diaz_Santos_2021}.

Also interesting to note is the prominent $9.7\mu\mathrm{m}$ silicate feature in our system. This matches with expectations from IR photometry data near the $9.7\mu\mathrm{m}$ feature in W2246-0526 \citep[and other Hot DOGs; see e.g.,][]{Tsai_2018, Li_2025}. Along some sightlines, the features are more or less prominent than others, depending on the geometry of the system at $r\lesssim 0.1~\mathrm{kpc}$ where most of the absorption occurs. Some of the diversity of the absorption lines might therefore be explained away as sightline variation. 

We stress that while this system does broadly align well with observed Hot DOG demographics, it also is broadly characteristic of the central galaxies of dense cosmological environments. Indeed, our zoom-in was originally chosen to be during the initial rapid growth period of an active SMBH phase in the most massive galaxy of a dense environment, prior to feedback. Perhaps what is most striking is that, despite this being the first simulation capable of reproducing the detailed near- to mid-IR obscuration required to generate a Hot DOG spectrum (since it is the first simulation to continuously resolve from the CGM to the dust torus), and despite the rarity of Hot DOGs observationally, we happen to catch the system in this state. This seems to suggests that Hot DOGs may be a key transitional period in the most massive galaxies of dense cosmological environments, at the beginning of an active phase, prior to outflows significantly clearing the system and turning the system into an optically-visible quasar. This picture also roughly agrees with expectations. If this indeed happens in an order one fraction of high-redshift super-Eddington quasars during its $\sim 10^{6-7}$ yr accretion period, then a typical $\sim 10^{4}$ km/s outflow will clear the region of obscuration ($\sim$10 pc) after $\sim 10^{3}$ yr, giving a Hot DOG event rate of one in every $\sim 10^{3-4}$ of such quasars, which roughly matches with the rates of high-redshift high-luminosity quasars discovered by WISE \citep{Eisenhardt_2012}. Indeed, considering Hot DOGs as a transition state prior to significant outflows has been suggested to explain Blue-Excess Hot DOGs \citep[BHDs;][]{Assef_2016, Assef_2020}, a subclass of Hot DOGs with excess UV emission, as a post-feedback stage where an outflow has cleared sufficient material to scatter optical emission from the AGN \citep{Finnerty_2020, Assef_2022, Li_2024, Liu_2025, Assef_2025}. To help determine if this is the case, we plan on zooming out from the black hole in future work, following the feedback outflows and its effect on the system's spectral classification and if it transforms into a BHD. Of course, this is only a case study of a single system, and future nuclear zoom-in simulations should reveal whether Hot DOGs are a common phenomenon in such massive galaxies or if this system is an outlier.

%%%%%%
\subsection{Starburst and AGN Component Decomposition}\label{sec:spectral_decomposition}

Since our simulation resolves from the galactic CGM to the dust torus, we can manually ``turn on/off" the AGN component of the system and study its effect on the spectrum. Figure \ref{fig:spectral_decomposition} shows our system decomposed into the stellar and stellar-heated dust component against the {emergent} AGN {(disk and torus after dust extinction)} and AGN-heated {(galactic)} dust components. The spectrum with the AGN component on appears like a Hot DOG, but when off (only stellar heating), it appears like a regular starburst galaxy. In the infrared, the AGN-heated dust dominates the mid-IR emission (from $\sim 2 \mu\mathrm{m}$ to $\sim 40 \mu\mathrm{m}$), and the starburst-heated dust dominates the near-IR emission ($\gtrsim 40 \mu\mathrm{m}$). The UV/optical is completely dominated by the stellar emission and none of the original AGN optical emission escapes to $r=100$ kpc. Since this system is high-redshift, metal-poor and young, the stellar UV/optical component is steep. {There is also a prominent stellar H$\alpha$ line, indicative of the rapid star formation occurring in this galaxy (rather than e.g., AGN narrow or broad lines).}

We also show the emergent spectrum out of the galactic nucleus ($r < 1$ kpc) and from the full galaxy merger system ($r < 100$ kpc) in Figure \ref{fig:spectral_decomposition}. The majority of the mid-IR emission is produced in the dense central nucleus, whereas the UV/optical and far-IR are produced primarily in the extended galaxy component. The UV/optical is perhaps most interesting since the majority of the escaping emission is produced by the extended galaxy component outside of the nucleus $r \gtrsim {1}$ kpc. Outside of the nucleus, the stellar emission is only slightly reddened, so the UV/optical spectrum still appears quite steep overall. The system therefore cleanly divides into three distinct dust regions: the dust torus (very dense gas, producing the majority of the hot dust component, as studied in \citealt{Bardati_2026}), the galactic nucleus that primarily reprocesses the dust torus emission, and the surrounding lightly reddened Lyman-break galaxy (producing the emerging optical/UV and far-IR). In section \ref{sec:hotdog_shape}, we investigate how these regions affect the resulting spectrum in more detail.

The deep silicate 9.7$\mu$m absorption features are predominantly produced in regions dominated by AGN emission, whereas the PAH features are found in regions powered by starburst. The majority of the silicate 9.7$\mu$m absorption feature is produced near to the dust torus, but sufficiently far to be in the surrounding ISM (outside of the black hole radius of influence) around $r\sim 100$ pc. We investigate the source of this in section \ref{sec:hotdog_shape}. Interestingly, there is a small amount extra emission in the combined spectrum that can be attributed to the starburst-heated PAH emission features (particularly the $7.7\mu\mathrm{m}$ feature). {This is most noticeable in the stellar-heated dust component of the system. We note that some of the less prominent lines (3.3 $\mu$m, 6.2 $\mu$m) are notably missing here due to the limitations of the wavelength grid used in the simulation.} It is possible that these PAH emission  may produce noticeable features in similar systems with slightly less luminous SMBH. Unfortunately, there currently no far-IR interferometers capable of resolving such rest-frame mid-IR dust features in Hot DOGs (particularly in high-redshift low-luminosity systems such as this one). 

Only at wavelengths $\lambda > 40\mu\mathrm{m}$ is the majority of emergent emission due to dust heated primarily by stars, with a significant contribution to the total emission out to $\lambda\sim 10^2 \mu\mathrm{m}$, particularly in the galactic nucleus. Even at wavelengths $\lambda > 150\mu\mathrm{m}$, $\sim 20\%$ of the dust emission is due to AGN-heating (reaching significantly into, e.g., Herschel bands). This is in contrast to the common observational assumption that the cold dust component can be attributed solely to stellar-heated dust. The effect would be even more noticeable in the total spectrum if the host galaxy were undergoing a weaker starburst or if the AGN were more luminous. {Indeed, estimating the star formation rate (SFR) using the far-IR luminosity $L_\mathrm{8-1000\mu\mathrm{m}}$ and the \citet{Bell_2003} calibration gives $\sim160 M_\odot/~\mathrm{yr}$, whereas excluding the AGN component in the luminosity yields $\sim110 M_\odot/~\mathrm{yr}$, which is closer to the actual SFR in this system. Thus, such far-IR estimates of SFR may be overestimates if there is a buried AGN providing some of the heating.} This point has been made using radiative post-processing of galaxy merger simulations that use {semi-analytic modeling or} sub-grid prescriptions black hole accretion and feedback {\citep{Schneider_2015, McKinney_2021, Di_Mascia_2021, Di_Mascia_2023}}, but here we confirm that this remains true in a system that resolves the AGN torus dust structure. {Interestingly, the AGN heating at these wavelengths occur due to absorption of the hot dust from the torus, rather than the disk emission. However, at later stages in the galaxy spectral evolution, it is possible that BH-driven outflows could generate optically-thin pathways that allow the disk emission to heat ISM dust directly}.

%While most of the optical emission is obscured by the dust torus and surrounding nucleus ISM, as we note in \citet{Bardati_2026}, dust clumping at unresolved (subgrid) scales may allow more AGN optical emission to escape than estimated here. Indeed, in \citet{Bardati_2026}, we even resolve some thin channels where light escapes at $r<1$ pc. At larger scales ($r\geq 10$ pc), there is not enough resolution in the simulation to resolve these channels if they exist. Including these channels could affect resulting optical spectrum and possibly leading to the system appearing as a Blue-Excess Hot DOG, or even optically visible quasar along these sightlines. 

%%%%%%%
\subsection{What Causes the Hot DOG's Spectral Shape?}\label{sec:hotdog_shape}

\begin{figure}[t]
    \centering
    \includegraphics[width=0.46\textwidth]{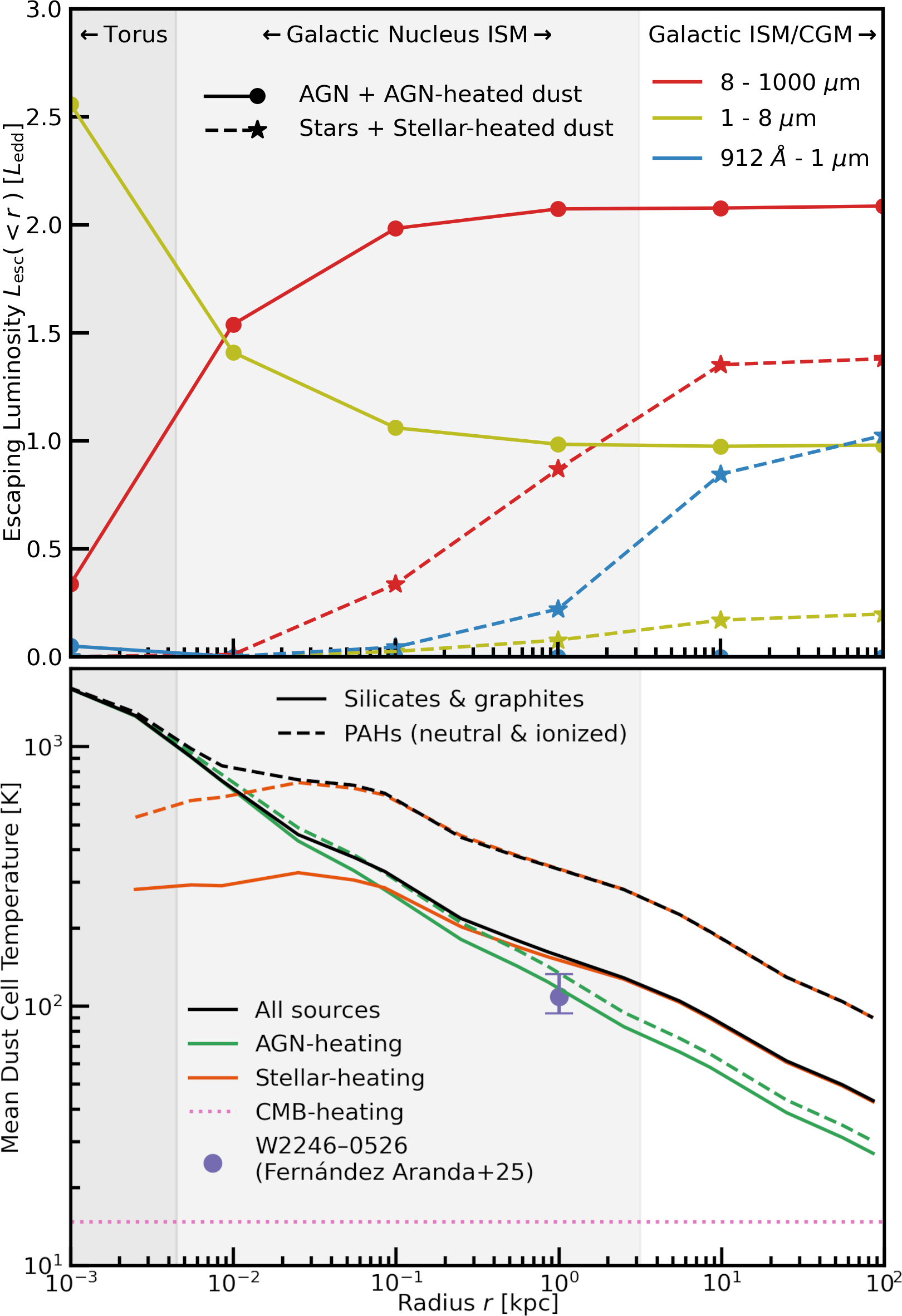}
    \caption{Radial profile of the escaping luminosity and dust temperatures, separated by the heating source. 
    \textit{Top panel:} We indicate the escaping luminosity of the {emergent} AGN {(disk and torus)} and AGN-powered dust with circles and star symbols denote the stellar and stellar-heated dust components. The emission is separated into the $8-1000~\mu\mathrm{m}$ mid- to far-IR emission, $1-8~\mu\mathrm{m}$ near- to mid-IR emission, and $912 \AA -1~\mu\mathrm{m}$ UV/optical to near-IR components. We also roughly mark the locations of the torus, the dense galactic nucleus and the more diffuse galactic ISM and CGM dust.
    \textit{Bottom panel:} We show the mean AGN-heated (green), starburst-heated (orange), and total (black) dust temperature profiles, separated into silicates \& graphites (solid lines) and stochastically-heated polycyclic aromatic hydrocarbons (PAHs, dotted lines). The cosmic microwave background (CMB) temperature at this redshift ($z=4.4$) is also indicated (purple dotted) for reference.
    The source of the reduced near-IR to mid-IR emission is from dust extinction from the galactic nucleus ISM. The galactic nucleus dust extinction also suppresses some of the nucleus optical emission of the starburst-heated dust, but the starburst also occurs outside of the dense nucleus ($r\sim 10$ kpc), producing significantly less reddened optical emission. Despite the luminosities being AGN-dominated at all scales, stellar heating dominates the mean dust temperatures past around $r \sim 100$ pc ($r \sim 10$ pc for PAHs) from nucleus dust extinction.}
    \label{fig:radial_luminosity_profile}
\end{figure}

A question that one could ask is exactly how and where in the galaxy is the Hot DOG spectrum actually produced. This is inherently a multi-scale problem since it requires resolving both the hot dust from the AGN dust torus and the warm/cold dust from the ISM/CGM, and is thus well suited to be answered by our simulation. In Figure \ref{fig:radial_luminosity_profile}, we show the radial profile of the escaping luminosity and the dust temperatures as a function of the heating source. We roughly denote the different spatial regimes in three parts: the dust torus (within a few parsecs), the dense ISM in the galactic nucleus (within a few kpc), and the optically thin galactic ISM. We address each regime in part.

As expected, the AGN-heated torus region broadly produces the hot dust emission ($T \gtrsim 10^3$ K). We discuss the structure and emission of the dust torus in this system in \citet{Bardati_2026}, and refer the reader to that paper for more details. Briefly, the torus is composed of a highly magnetized disk and dusty tidal stream fueling accretion. Although it has a clear circularized ``torus" composed of the outer accretion disk, as shown in Figure \ref{fig:radial_luminosity_profile}, the torus region is almost entirely spherically optically thick due to the extreme dust density in the region. We show in \citet{Bardati_2026} that these extreme densities can cause azimuthally asymmetric near- to mid-IR from significant dust self-absorption in non-axisymmetric structures such as the tidal stream fueling accretion. The emergent luminosity from the torus region is thus mostly hot dust that is progressively reprocessed to cooler temperatures, with little to no UV/optical escape from the AGN.

Progressing well outside of the black hole radius of influence and into the inner galactic nucleus ($r \sim 10 - 10^2$ pc), the dust (composed overall mostly by silicates and graphites) is still heated primarily by the AGN. In this region, however, the PAH temperatures decouple from the rest and are heated significantly higher mean temperatures because of stochastic stellar heating of small grains. The PAH heating peaks around $r\sim 30$ pc, which roughly corresponds to the starburst peak in the galactic nucleus. Nearly all of the stellar UV/optical emitted in the inner nucleus is dust-absorbed, beginning to produce significant stellar-heated warm dust ($T \lesssim 300$ K, $\lambda \gtrsim 10~\mu\mathrm{m}$), which now mostly passes unattenuated to the observer. However, the majority of the warm dust still comes from dust absorbed re-emission of the hot dust torus emission. The significant absorption in this region explains the sharp cutoff in the near- to mid-IR around $\lambda \lesssim 5\mu\mathrm{m}$. Since the degree of nuclear obscuration can greatly vary from galaxy to galaxy, this may help explain some of the variation in the $2-8\mu\mathrm{m}$ shape of Hot DOGs. {This seems to match with the recent ALMA observations of W2305-0039 by \citet{Tadaki_2026} that show evidence of a deeply buried nucleus with rising CO(11-10)/CO(7-6) ratios up to $\sim500$ pc from a super-Eddington accreting SMBH.}

In the outer galactic nucleus ($r \sim 0.1 - 1$ kpc), the dust is now dominated by stellar heating (even the silicates and graphites). The optical depths are now low enough that dust self-absorption is minimal and thus the AGN-heated IR luminosity is mostly constant, only now affecting the far-IR, cold dust tail (which is dominated by stellar emission in the emerging $r>100$ kpc spectrum anyway). In this region, stellar-heated dust begins to contribute meaningfully to the total mid- to far-IR emission. Moreover, stellar UV/optical emission begins to escape, but it is significantly reddened and will be dominated by stellar emission at larger radii. 

In the extended galactic and circumgalactic region ($r > 1$ kpc), the optical depth drops substantially, allowing for UV/optical emission from the young stellar population with only some reddening (as seen in Figure \ref{fig:spectral_decomposition}). The dust in this region is {mostly} stellar-heated, but there is little dust absorption to substantially affect the IR luminosity (other than the far-IR tail). Overall, the spectral shape of our system is thus broadly generated from 1) AGN-heated hot dust emission produced in the torus that is reprocessed down to colder temperatures in the inner galactic nucleus ISM yielding the reduced $3-8~\mu\mathrm{m}$ emergent luminosity in our spectra, 2) the mostly starburst-heated far-IR tail, generated primarily outside of the galactic nucleus, and 3) the lightly reddened young stellar population in the extended galaxy that produces the steep UV/optical spectrum.

It is worth noting that the depth of the $9.7~\mu\mathrm{m}$ silicate feature stabilizes when the dust self-absorption begins to significantly decline $r\gtrsim 100$ pc, which is well outside of the region commonly associated with the dust torus. This nicely illustrates that, at least in this highly optically thick galactic nucleus, this silicate feature is not directly produced by the dust torus, as is sometimes assumed, but rather the galactic nucleus ISM. Indeed, observational data of Compton-thick AGN have suggested that these features are mostly generated in the nucleus ISM \citep{Goulding_2012}. {Typically, however, radiative transfer torus models assume that this silicate feature is a part of the torus and tend to find deep silicate features in large, optically thick, smooth tori, often used as evidence for a clumpy torus \citep{Feltre_2012}. Naively, it is at least possible that such models may fit part of the diffuse ISM, rather than the torus structure itself. In future work, we plan on investigating this further by fitting such torus models back onto these synthetic spectra and comparing the inferred torus geometries.} %The PAH emission features are instead predominantly generated in the extended galaxy ($r \gtrsim 1$ kpc). Since the PAH temperature decreases as a function of radius, producing a decreasing PAH ionization fraction (emerging $7.7~\mu\mathrm{m}/11.3~\mu\mathrm{m}$ flux ratio is $\sim 1$ at 100 pc, but $\gtrsim 1.3$ at 100 kpc), the resulting PAH ionization fraction inferred from the spectrum would be higher than it actually is in the optically thick molecular environment where the starburst primarily occurs. 

The temperature profile is remarkably consistent with the recent work of \citet{Fernandez_Aranda_2025}, which produced spatially resolved temperature maps for W2246–0526 using ALMA observations. They find a dust temperature of $T_d \sim 110~\mathrm{K}$ at a nucleus radius $r\sim 1$ kpc (similar to ours), which they attribute to AGN-heated dust. At larger radii, $r > 10$ kpc, the W2246–0526 temperature map drops to $\sim 30-40~\mathrm{K}$, which they attribute to stellar-heated dust. Their nucleus temperatures are strikingly similar to the AGN temperature at similar radii in our system, despite the two order of magnitude difference in luminosity between our system and W2246–0526. If we account for the factor of $\sim 10^2$ in luminosity between the systems with a $\sim 10$ factor decrease in radius (following a dust thermal equilibrium relation $R\propto L^{0.5}$, a typical scaling for the broad line region/inner dust torus radius), then the W2246–0526 datapoint would actually match well to the stellar-heated line in Figure \ref{fig:radial_luminosity_profile}. Of course, more equivalent systems are needed to make any substantive claims comparing Hot DOG temperature mapping observations and nuclear zoom-in simulation predictions, but it is interesting to note the similarity nonetheless.

\subsection{The Low-Luminosity Nature of this Hot DOG}

Despite strongly resembling Hot DOGs in shape and general system properties, our simulation would actually not be detected in the WISE survey primarily due to its lower luminosity than most W1W2 dropouts at similar redshifts ($z = 4.4$). Our quasar system has an IR luminosity of $L_\mathrm{IR} \approx 7 \times 10^{45}~\mathrm{erg~s^{-1}} \approx 2 \times 10^{12} L_\odot$, compared to the typical $\sim 10^{13 - 14} L_\odot$ luminosities of WISE Hot DOGs \citep{Tsai_2015}. The absence of such lower-luminosity Hot DOGs in observations is likely simply a selection effect from the flux limits of the WISE survey. Indeed, detections of relatively local Hot DOG systems show similar IR luminosities and shapes, but with older stellar populations \citep{Li_2023, Li_2025}. Our work thus seems to suggest that a class of lower-luminosity Hot DOGs could exist at high-redshifts. 

Fundamentally, the luminosity differences between our system and high-redshift Hot DOG observations are likely due to differences in black hole masses of a similar factor ($1.3\times10^7M_\odot$ vs $\sim 10^{8-9}M_\odot$). Our system zoom-in was chosen to be during an intense quasar-fueling event, in a galaxy that was a rapidly growing outlier among a sample of the progenitors of $\sim 10^{13}$ dark matter halos at $z=0$. However, most systems similar to the one we study in this paper (at $z=4.4$) would eventually continue to grow to be much larger than $\sim 10^{13}$ at $z=0$, possibly to become cluster-sized, thus observationally classifying this system at $z=4.4$ as a proto-cluster. To obtain a higher luminosity Hot DOG relative to its host galaxy, one could choose a similar system that rapidly accretes for longer without clearing winds, undergoes multiple rapid accretion events, has a larger initial black hole mass, or has a relatively more typical galaxy growth evolution. Indeed, many of the observed high-luminosity Hot DOGs lie above the $M$-$\sigma$ relation, have outflow components, and have been suggested to result from multiple accretion periods driven by frequent in-fall of neighbouring galaxies and clumps \citep{Finnerty_2020, Diaz_Santos_2021, Li_2024, Liu_2025, Assef_2025}. 

To detect these high-redshift low-luminosity Hot DOGs, future infrared survey telescopes such as the Probe Far-Infrared Mission for Astrophysics (PRIMA) would be instead required. Our system in particular is on the edge of what can be detected with the PRIMA imager, requiring $\gtrsim 10^2$ hours of exposure time in a 10 square degree survey area to reach $> 5\sigma$. However, we argue that this will depend strongly on the rarity of such low-luminosity systems and that similar systems with luminosities in between the WISE detected high-redshift Hot DOGs and ours should still be easily detectable. After such detections, follow-ups with near-IR and mid-IR spectroscopy, such as the James Webb Space Telescope's (JWST) NIRspec or MIRI should be able to detect the rest-frame UV/optical and near-IR emission in galaxies with young stellar populations similar to our simulated system.

\section{Conclusions}\label{sec:conclusion}

In this paper, we analyze a FORGE'd in FIRE simulation zoomed in from cosmological scales to the accretion disk of a super-Eddington accreting $\sim 1.3 \times 10^7 M_\odot$ supermassive black hole in the center of a starburst galaxy merger, prior to supermassive black hole (SMBH) feedback \citep[described in][]{Hopkins_2024a}. The simulation includes non-ideal magneto-hydrodynamics, multi-band radiation-hydrodynamics coupled to multi-species thermochemistry and individual (proto)star and stellar population formation and feedback, and SMBH seeding and accretion. 

Since this simulation continuously resolves both galaxy interstellar medium scales and dust torus scales, it can uniquely be used to generate a self-consistent near- to far-infrared spectrum of the starburst quasar system, without the use of any phenomenological torus models. This allowed us in \citet{Bardati_2026} to study the structure and emission signatures of this system's dust torus, including showing that the dust torus is composed largely of the highly magnetized accretion disk and is azimuthally anisotropic in the mid-infrared due to significant dust self-absorption (such as from a tail of dense dusty gas feeding the accretion). In this paper, we extend this radiative analysis out to galactic scales, generating the first self-consistent emergent galactic infrared spectrum of a highly star forming and obscured quasar system. We find the following:

\begin{enumerate}
    \item \textbf{The system has a broadly flat spectral index infrared spectrum, similar to a Hot Dust Obscured Galaxy (Hot DOG).} Our system also matches with many other observational properties of Hot DOGs such as undergoing multiple galaxy mergers, being Compton-thick and morphologically clumpy, having prominent $9.7~\mu\mathrm{m}$ silicate absorption features, and existing in the massive central galaxy of a high-redshift, dense cosmological environment (in what would be classified as a proto-cluster). Our system is in the early stages of a super-Eddington accretion period before significant SMBH feedback and outflows.
    \item \textbf{Our system cleanly divides into three regions: 1. the dust torus (as studied in \citealt{Bardati_2026}), 2. an optically-thick galactic nucleus (that reprocesses the torus and starburst emission), 3. an optically-thin extended ISM and young starburst galaxy.} The reduced $2 - 8 \mu\mathrm{m}$ emission characteristic of Hot DOG spectra is due to substantial dust self-absorption by the surrounding dense ISM dust in the galactic of our system. This self-absorption mostly occurs around $1 - 100$ pc, well outside the dust torus region. This self-absorption can vary across sightlines, leading to variation in the near-IR to mid-IR spectra. The starburst produces the majority of the overall dust heating outside of the self-absorption region ($r \gtrsim 100$ pc), but dominates PAH heating significantly closer ($r \gtrsim 10$ pc) from stochastic heating of small grains.  % sentence moved from below
    \item \textbf{The mid-IR is dominated by the AGN component, and the far-IR is mostly produced by the starburst component but still has a significant AGN contribution to the cold dust emission ($\mathbf{\sim 20\%}$ of observed spectral luminosity), particularly in the galactic nucleus.} {This causes an overestimation of the galactic star formation rate of $\sim 50\%$, when using the 8-1000 $\mu\mathrm{m}$ luminosity as a tracer compared to the true value in this synthetic galaxy.}
    \item {\textbf{The system has a prominent $\bm{9.7 \mu\mathrm{m}}$ silicate absorption line, produced primarily in the surrounding optically thick ISM dust, not the dust torus itself.} Since radiative transfer dust torus models often assume the opposite, this may have implications for such modeling in similar galaxies with dense nuclear regions. } 
    %\item \textbf{Due to sightline dependence of the highly optically thick and azithuthally asymmetric mid-IR dust torus, a broad range of IR spectral indexes can be inferred, even in the same intrinsic galaxy.} Naively, this can transform the near-IR of a system greatly .... Notably, the $9.7$ varies greatly across sightline (by at about an order of magnitude), potentially occasionally affecting its use as a diagnostic of AGN-driven dust emission.
    \item \textbf{The system is lower luminosity than similar redshift ($\mathbf{z\approx 4.4}$) Hot DOGs observed by WISE likely due to selection effects, but may be detectable by future mid- to far-IR surveys such as PRIMA.} While this particular system is near the boundary of what is possible to detect at $\geq5\sigma$, PRIMA should easily detect the bridge between high luminosity Hot DOGs (detected by WISE) and lower luminosity analogs such as this one. Targeted near- to mid-IR follow-ups of such detections (e.g. JWST) could then reveal the rest-frame optical-MIR and host galaxy emission.
\end{enumerate}

To our knowledge, this work represents the first self-consistently simulated hot dust obscured galaxy spectrum. Broadly, our results suggest that hot dust obscured galaxies may result as a super-Eddington accreting SMBH phase in dusty high-redshift starbursts due to mergers, prior to significant BH feedback. This seems to match with observations of high-luminosity Hot DOGs. This paper is only a case study though, and the question of the nature of Hot DOGs would ultimately be best answered with a sample of high-redshift SMBH zoom-ins. Since our system is beyond the flux and redshift limitations of WISE, we hope that this work will inform future mid to far-infrared telescope surveys in searching for low-luminosity and high-redshift ($z>4$) versions of this rare galaxy type.

After having now refined this FORGE'd in FIRE simulation further to resolve the innermost stable orbit and resulting black hole feedback \citep[described in][]{Hopkins_2025}, we plan on evolving the simulation back out to galactic scales to see how the black hole-driven outflows affect the surrounding nuclear and galactic environment. This will be useful for a variety of key science goals, including understanding the effect of SMBH feedback on the system and the spectral evolution of this ultraluminous infrared galaxy. It has been suggested that outflows could clear an optically thin pathway, transitioning the galaxy from a warm ULIRG to a red quasar and eventually a blue quasar. Such future work could help constrain this process and explain the observational correlation between warm ULIRG spectra and the existence of galactic molecular outflows.  

%%%%%%%%%%%%%%%%%%%%%%%%%%%%%%%%%%%%%%
\begin{acknowledgments}
{The authors thank the anonymous referee for their detailed comments}. The authors {also} thank {Peter Eisenhardt,} Gordon Richards, {Roberto Assef, Ken Tadaki, Rom\'{a}n Fern\'{a}ndez Aranda, Daniel Stern,} Adam Leroy, Sam Ponnada, and Haiyang Wang for insightful comments and discussions. The numerical calculations of this work were run on the Frontera supercomputer with Texas Advance Computing Center (TACC) allocation AST21010. J.B. acknowledges support from a doctoral scholarship provided by the Natural Sciences and Engineering Research Council of Canada (NSERC). P.F.H. acknowledges support from a Simons Investigator Grant. C.{-}A.F.{-}G. was supported by NSF through grants AST-2108230 and AST-2307327; by NASA through grants 80NSSC22k0809, 80NSSC22K1124 and 80NSSC24K1224; by STScI through grant JWST-AR-03252.001-A; and by BSF through grant \#2024262.
\end{acknowledgments}

%%%%%%%%%%%%%%%%%%%%%%%%%%%%%%%%%%%%%%

\bibliography{paper}{}
\bibliographystyle{aasjournal}

\end{document}